\documentclass[12pt]{aastex}
\usepackage{natbib}
\usepackage{graphicx}
\usepackage{epsf,amsfonts,amsmath,amssymb}
\usepackage{mathrsfs}
\usepackage[usenames,dvipsnames,svgnames,table]{xcolor}
\usepackage{ulem,color}
\usepackage{array}
\usepackage{epstopdf}
\usepackage{multirow}
\usepackage{ulem}
\usepackage{rotating}

\begin{document}
\author{Paz Beniamini, Kenta Hotokezaka and Tsvi Piran}
\affil{ Racah Institute of Physics, The Hebrew University of Jerusalem, Jerusalem 91904, Israel
}

\title{r-process Production Sites as inferred from Eu Abundances in Dwarf Galaxies}

\begin{abstract}
{Recent observations of $r$-process material in ultra-faint dwarf galaxies (UFDs) shed light on the sources of these elements.
Strong upper limits on the Eu mass in some UFDs combined with detections of much larger masses in a UFD, Reticulum II, and other dwarf galaxies
imply that Eu production is dominated by rare events, and that the minimal Eu mass observed in 
any UFD is approximately the amount of Eu mass produced per event. This is consistent with other independent observations in the Galaxy. 
We estimate, using a model independent likelihood analysis, the  rate and   Eu (Fe)  mass produced per $r$-process (Fe production) event in dwarf galaxies
including classical dwarfs and UFDs. The mass and rate of the Fe production events are consistent with 
the normal  core-collapse supernova~(ccSN)   scenario.
The Eu mass per event is  $3\times 10^{-5}M_{\odot}<\tilde{m}_{\rm Eu}<2\times 10^{-4}M_{\odot}$, corresponding to a total $r$-process mass per event of $6\times 10^{-3}M_{\odot}<\tilde{m}_{r-process}<4\times 10^{-2}M_{\odot}$. The rate of $r$-process events is $2.5\times 10^{-4}<R_{rp/SN}<1.4\times 10^{-3}$ as compared with the ccSNe rate.
These values are consistent with the total Eu mass observed in our own Galaxy, suggesting that the same  mechanism
is behind the production of $r$-process events in both dwarf galaxies and the Milky Way, and that it may be the dominant mechanism for production of $r$-process elements in the Universe.
The results are consistent with neutron star mergers estimates, but cannot rule out other rare core collapse scenarios, provided that they produce a significant amounts of 
$r$-process material per event. }

\end{abstract}

\section{Introduction}
\label{sec:Introduction}
The origin of about half of the elements heavier than Fe, produced via the rapid neutron capture process~($r$-process), is still a mystery (\citealt{cowan1991PhR,wanajo2006NuPhA,qian2007PhR,arnould2007PhR}; and see \citealt{thielemann17} for a recent review).
Core-collapse supernovae~(ccSNe) have long been thought of as the production sites 
\citep{burbidge1957RvMP,cameron1957}. This scenario
requires either a neutrino driven wind with high entropy~\citep{woosley1994ApJ, qian1996ApJ,hoffman1997ApJ,goriely2015},
a highly magnetized outflow~\citep{suzuki2005ApJ,metzger2008ApJ,Winteler12,vlasov2014MNRAS,nishimura2015ApJ,Moesta15}, 
wind from a black-hole torus~\citep{wanajo2012ApJ}, or fall-back supernovae~\citep{fryer2006ApJ}.
It is unclear whether any one of those occurs frequently enough \citep{Qian2000ApJ}.
With this realization the alternative scenario of 
compact binary mergers (neutron star--neutron star or neutron star--black hole binaries) 
that eject highly neutron rich material in which $r$-process nucleosynthesis takes place
\citep{lattimer1976ApJ,symbalisty1982ApL,eichler1989Nature,freiburghaus1999ApJ},
has been attracting more and more attention.
Nucleosynthesis studies have shown that all $r$-process nuclides
can be produced in the neutron star merger outflows~\citep{wanajo2014ApJ,wu2016}.
The discovery of macronova/kilonova candidates after short gamma-ray bursts~\citep{tanvir2013Nature,berger2013ApJ,yang2015NatCo,jin2016}
provided recent support to this idea.

Dwarf galaxies are composed of old stellar populations~(e.g. \citealt{weisz2015ApJ}).
The chemical abundances of  dwarf galaxies has been frozen since shortly after 
those galaxies were formed. These chemical abundances have  unique implications for the $r$-process production scenario~(e.g. \citealt{tsujimoto2014A&A}).
$r$-process elements have been detected in  classical dwarf galaxies
\citep{shetrone2001ApJ,shetrone2003AJ, venn2004AJ,letarte2010A&A, tsujimoto2015PASJ}.
In addition to stable $r$-process elements, thorium is found in a star of Ursa Minor and its abundance ratio to stable 
$r$-process elements is lower than solar \citep{aoki2007PASJ}, suggesting that 
events producing heavy $r$-process elements took place in this galaxy
in the early Universe $\sim 10$~Gyr ago.

The {\it Sloan Digital Sky Survey} 
discovered several ultra-faint dwarf galaxies~(UFDs; e.g., \citealt{willman2005ApJ,belokurov2007ApJ}).
Remarkably, these UFDs contain only $\sim 10^{3}$ to $10^{5}$ stars and
they are composed of an old and extremely metal poor stellar population,
suggesting that the UFDs are the closest objects to the first galaxies~(see \citealt{bromm2011} for a review).
For some of them, the majority of the stars were likely to have formed before reionization, corresponding to a Universe age 
of $\sim 13$~Gyr old~\citep{brown2014ApJ}. 
Recently, \cite{ji2016Nature} and \cite{roederer2016AJ} reported the first discovery of highly $r$-process enriched stars
in a UFD, Reticulum II, while there are only strong upper limits on the abundances of $r$-process elements
for some other UFDs~\citep{frebel2010ApJ,frebel2014ApJ,roederer2014MNRAS}. 
These measurements that show large fluctuations in the $r$-process abundance imply that only a single $r$-process event took place in Reticulum II. 
Following these observations of $r$-process material in the classical dwarfs and UFDs,
a question that naturally arises is what are the rate and amount of matter produced in $r$-process events that are consistent with these observations.

We estimate here the most likely mass ejection of Fe and Eu and rate  of ccSNe per total luminosity 
 and of $r$-process production events relative to ccSNe. We
use a likelihood analysis, based on Poisson statistics. The formulation is independent of the specific nature of the sources of $r$-process 
elements. We use the given measured Fe and Eu abundances in dwarf galaxies and their initial gas masses (to which those metals were injected). We assume that all dwarf galaxies (or alternatively, only UFDs) share the same mechanism for $r$-process production. The results in both cases (of all dwarf galaxies and just UFDs) strongly suggest that the $r$-process event rate is considerably smaller than the ccSN rate. They are comparable to the expectations from the neutron star merger scenario.

This paper is organized as follows. In \S \ref{sec:sample} we describe the sample of UFD galaxies.
A critical factor in the analysis is the initial gas masses in UFDs.  
We estimate the initial gas masses in UFDs in \S \ref{sec:initial} 
using several different methods. We turn to the initial masses of Fe and $r$-process materials in \S \ref{sec:initialRproc}. 
We apply a likelihood analysis in order to examine the Fe and $r$-process abundances in dwarf galaxies
and estimate the rates and ejecta mass of ccSNe and $r$-process events needed in \S \ref{sec:like}. 
We then compare the implied rate and $r$-process yields with those expected in the double neutron star~(DNS) merger scenario in \S \ref{sec:DNSRates}
and peculiar ccSNe in \S \ref{sec:ccSNe}. Finally, we summarize our results in \S \ref{sec:conclusions}.

\section{The galaxy sample}
\label{sec:sample}
We consider dwarf spheroidal galaxies and focus in particular on UFDs that are
composed of old and metal poor stars~(e.g. \citealt{simon2007ApJ}). 
We consider galaxies for which there are reported detections or upper limits on the Fe and Eu abundances.
Our sample includes five UFDs: Reticulum II, Segue I, Segue II, Coma Berenices and Ursa Major II.
In addition we consider 6 more massive classical dwarf galaxies: Carina, Sculptor, Draco, Leo I, Ursa Minor 
and Fornax. We do not consider Sagitarius, since this galaxy has been shown to be considerably tidally disrupted \citep{ibata1995MNRAS,mateo1996ApJ,majewski2003ApJ}, and its mass is
therefore very uncertain. Table~\ref{tbl:data} shows
the V-band luminosities, $L_V$, 
the velocity dispersions, $\sigma$, and the half light radii, $r_{1/2}$ taken from \cite{walker2009ApJ,walker2015ApJ},
and the average Fe and Eu abundances~\citep{shetrone2001ApJ,shetrone2003AJ,frebel2010ApJ,letarte2010A&A,frebel2014ApJ,
ji2016Nature,ji2016,roederer2014MNRAS,tsujimoto2015PASJ, roederer2016AJ}.
The errors of the average abundances shown in Table~\ref{tbl:data} include statistical and systematic uncertainties.
If the systematic errors of the measurements for a galaxy are not available in the literature, 
we simply assume a systematic error of $0.2$~dex.

\begin{table*}[ht]
\begin{center}
%\begin{minipage}{18cm}
%\multicolumn{1}{|c|}{\centering $M_{h}(<r_{1/2})$ \\ $[10^{7}M_{\odot}]$} 
\caption[]{
  Summary of observed properties of dwarf galaxies.}
%\hspace{-2.2cm}
\scalebox{0.75}{
 \begin{tabular}{cccccccc} \hline \hline
Type & Object & $r_{1/2}$~[pc] & $\sigma~[{\rm km/s}]$ & $M_{h}(\!<\!r_{1/2}) [10^{7}M_{\odot}]$ & $L_{V}~[L_{V,\odot}]$ & ${\rm \langle [Fe/H]\rangle}$ & ${\rm \langle [Eu/H]\rangle}$  \\ \hline 
\multirow{5}{*}{UFDs} & Reticulum II & $32^{+1.9}_{-1.1}$ & $3.6_{-0.7}^{+1.0}$   & $0.024^{+0.014}_{-0.008}$& $1.0\pm 0.09\cdot 10^3  $& 
$-2.83\pm0.23$ & $-0.97\pm0.19$\\
& Segue I & $29\pm7$ & $4.3\pm 1.2$ & $0.031\pm 0.019$ & $3.3\pm2.1\cdot 10^2$ & $-2.68\pm 0.21$ & $<-2.0$  \\
& Segue II & $34\pm5$& $3.4\pm 1.8$ & $0.023\pm 0.023$ & $8.5\pm1.7\cdot 10^2$ & $-2.96\pm0.19$ & $<-3.0$ \\
& Ursa Mayjor II   & $140\pm25$ & $6.7\pm 1.4$ & $0.36\pm0.16$  & $4.0\pm1.9\cdot 10^3$ & $-2.89\pm 0.22$ & $<-3.1$ \\
& Coma Berenices& $77\pm10$  & $4.6\pm 0.8$ &  $0.094\pm 0.035$&  $3.7\pm1.7\cdot 10^3$& $-2.57\pm 0.22$ & $<-2.4$ \\ 
\hline
\multirow{6}{*}{classical dwarfs} 
& Draco    & $196\pm12$ & $9.1\pm 1.2$   &  $0.94\pm 0.25$ & $2.7\pm0.4\cdot 10^5$ & $-2.00\pm 0.21$ & $-1.26\pm 0.26$ \\
& Leo I     & $246\pm19$  & $9.2\pm 1.4$   &  $1.2\pm 0.4$  & $3.4\pm1.1\cdot 10^6$ & $-1.29\pm 0.20$ & $-0.75\pm 0.24$  \\
& Fornax & $668\pm34$  & $11.7\pm 0.9$  & $5.3\pm 0.9$  & $1.4\pm0.4\cdot 10^7$ & $-0.90\pm 0.15$ & $-0.33\pm 0.22$\\
%Sgr & $1550\pm50$ & $5800\pm 190$ & $11.4\pm 0.7$ & $12\pm1.0$ & $81\pm 13$ & - \\
& Carina  & $241\pm23$  & $6.6\pm 1.2$   & $0.61\pm0.23$ & $2.4\pm1.0\cdot 10^5$  & $-1.64\pm 0.20$ & $-1.34\pm 0.21$ \\
& Ursa Minor  & $280\pm39$ & $9.5\pm 1.2$   & $1.5\pm0.4$   &  $2.0\pm0.9\cdot 10^5$ & $-1.90\pm 0.11$& $-1.11\pm 0.23$\\
& Sculptor & $260\pm39$& $9.2\pm 1.1$   & $1.3\pm 0.4$ &  $1.4\pm0.6\cdot 10^6$  & $-1.64\pm 0.20$ & $-1.16\pm 0.22$ \\
\hline 
\label{tbl:data}
\end{tabular}
}
\end{center}
%\end{minipage}
\end{table*}
\section{The chemical evolution of dwarf galaxies}
\label{sec:initial}
{To evaluate the number of events that supplied Eu or Fe in each galaxy, 
the total amount of the elements produced in it is needed.
We assume that the elemental abundances
of stars in dwarf galaxies represent the amount of the elements in the gas from which the stars have been formed.
The total amount of Fe and Eu produced in a galaxy during active star formation is
estimated using the elemental abundances listed in Table \ref{tbl:data}, together with the gas mass
into which metals are injected. The relatively low stellar luminosities as compared with the total halo masses 
suggest that UFDs  have lost most of their initial gas 
due to either supernova feedback or IGM reionization~\citep{bovill2009ApJ}.}
Thus the observed upper limits ($\sim 10^{3}-10^{4}M_{\odot}$) on their current gas mass are not representative of the amount of material into which the $r$-process elements have been  injected at the early stages of their evolution.
Therefore, we estimate the amount of the gas  based on the assumption that the initial gas mass 
is proportional to the halo mass within the size of the stellar spheroidal.

When considering the mass loss from the dwarf galaxies we assume that the gas is chemically homogeneous with the average
abundances, as observed in the remaining stars. However, of course, the chemical mixing process in the gas
is inhomogeneous and a fraction of the produced material may escape 
from the galaxies, or alternatively, gas with a primordial composition may escape. Such effects should be taken into account when
considering the abundance distribution of stars in each dwarf galaxy. 
These issues should be addressed with hydrodynamics simulations~(see e.g. \citealt{pallottini2014MNRAS,ritter2015MNRAS,hirai2015}) and are beyond
the scope of this work. 

The initial gas mass, $M_g$, should be 
between the total stellar mass of the system and $1/6$ of the total halo mass of the system
(we take here the typical fraction of $1/6$ for the ratio of Baryonic to total mass in a halo). 
As mentioned above, however, the stellar mass of UFDs is significantly smaller than 
the halo mass due to significant gas loss throughout the formation and evolution of the galaxy.
We estimate $M_{g}$ using the halo mass within a given radius, $M_{h}(<r)$.
This can be calculated directly from measurements of the half-light radius $r_{1/2}$ and the velocity dispersion $\sigma$
by using the Jeans equation~(e.g. \citealt{walker2009ApJ}):
\begin{eqnarray}
M_{h}(r) = \frac{5r_{1/2}\sigma^2(r/r_{1/2})^3}{G[1+(r/r_{1/2})^2]}\approx  1.2\times 10^{3} M_{\odot} 
\left(\frac{r}{1~{\rm pc}} \right) \left(\frac{\sigma}{1~{\rm km/s}} \right)^2\frac{(r/r_{1/2})^2}{1+(r/r_{1/2})^2},\label{mass} 
\end{eqnarray}
where the Plummer model for the stellar distribution with an isotropic and constant velocity dispersion is assumed. 
The halo masses within the half-light radius estimated with this formula by $M_{h}(<r_{1/2})=5r_{1/2}\sigma^2/2G$ are
shown in Table~\ref{tbl:data}. 
Of course, this formula can only be used up to the radius where the above assumptions are valid.
As a fiducial model, we use $M_g\approx M_h(<r_{90})/6$,
where $r_{90}\approx 3.71r_{1/2}$. According to the Plummer distribution, this choice corresponds to assuming that
the material within the radius $r_{90}$, where $90\%$ of the stellar mass is included, contributes to the chemical
mixing of the injected metals.  
For comparison, we consider also initial halo masses that are smaller, $M_h(<r_{50})$, or lager
than $M_h(<r_{90})$ by a factor of 3.

As the assumptions made in Eq.~(\ref{mass}) may be invalid at $r_{90}$ for some large dwarf galaxies,
we also employ an alternative way to estimate the halo masses within $r_{90}$:
using the Navarro-Frenk-White (NFW) mass profile with the model parameters derived 
in \cite{walker2009ApJ}, and integrating the distribution up to $r_{90}$.
This method can only be carried out for the classical dwarfs.
For the UFDs there are insufficient observations of stellar kinematics at large enough radii that would constrain the model parameters.
In all but one case (Fornax), this estimate results in similar values to those obtained with the first method.
Table ~\ref{tbl:mass} provides a summary of the different masses.
% as well as the minimum escape velocities.
% (see Appendix for discussions of escape velocities).

We compare the gas masses estimated here with those from the literature.
Using a cosmological simulation of the 
first galaxies, \cite{greif2010ApJ} show that Population II stars are formed
from cold gas of $\sim 10^{5}M_{\odot}$  at the center of a galaxy with a total 
mass of $\sim 10^8M_{\odot}$. The smallest gas masses in our sample,
$2$--$3\cdot 10^5 M_{\odot}$ for Reticulum II, Segue I, and  Sugue II,
are consistent with the result of \cite{greif2010ApJ}.

\section{$r$-process elements in dwarf galaxies}
\label{sec:initialRproc}

The total masses of Fe and Eu produced in the system are estimated as
\begin{eqnarray}
M_{\rm Fe} & \approx &120\cdot 10^{[\rm{Fe/H}]} M_{\odot}   \left(\frac{M_{g}}{10^{5}M_{\odot}}\right),\label{eu}\\
M_{\rm Eu}  & \approx  &3.7\cdot 10^{-5+[\rm{Eu/H}]} M_{\odot}   \left(\frac{M_{g}}{10^{5}M_{\odot}}\right),\label{eu}
\end{eqnarray}
where we assume that metals with an Eu (Fe) abundance of [Eu/H] ([Fe/H]) are injected into a gas with an initial mass $M_{g}$ and 
used the solar abundances from \cite{lodders2003ApJ}.

While only the abundances of  Eu and some other $r$-process elements are measured,
the total mass of $r$-process elements produced in a dwarf galaxy can be estimated 
by extrapolating the Eu abundance to other $r$-process ones using the solar $r$-process abundance pattern~\citep{goriely1999}:
\begin{eqnarray}
M_r \sim \begin{cases}
3.3\cdot ~10^{-2+[\rm{Eu/H}]} M_{\odot}   \left(\frac{M_{g}}{10^{5}M_{\odot}}\right)~~~~~(A_{\rm min}=70),\\
0.7\cdot ~10^{-2+[\rm{Eu/H}]} M_{\odot}   \left(\frac{M_{g}}{10^{5}M_{\odot}}\right)~~~~~(A_{\rm min}=90), \label{rmass}\\
0.6\cdot ~10^{-2+[\rm{Eu/H}]} M_{\odot}   \left(\frac{M_{g}}{10^{5}M_{\odot}}\right)~~~~~(A_{\rm min}=110),
\end{cases}
\end{eqnarray}
where $A_{\rm min}$ is the minimum mass number of $r$-process elements.
It is worthwhile noting that the abundances of the $r$-process elements 
with atomic numbers of $Z<40$, corresponding to $A<90$, in Reticulum II 
are depleted compared to the solar~\citep{ji2016} and similar to those of
an extreme metal-poor star CS~22892-052 \citep{sneden2003ApJ}. This suggests that there are two types
of events, light and heavy $r$-process events, and only the latter took place
in Reticulum II.

\begin{figure}[h]
\includegraphics[angle = 270, bb=50 50 554 770,width=90mm]{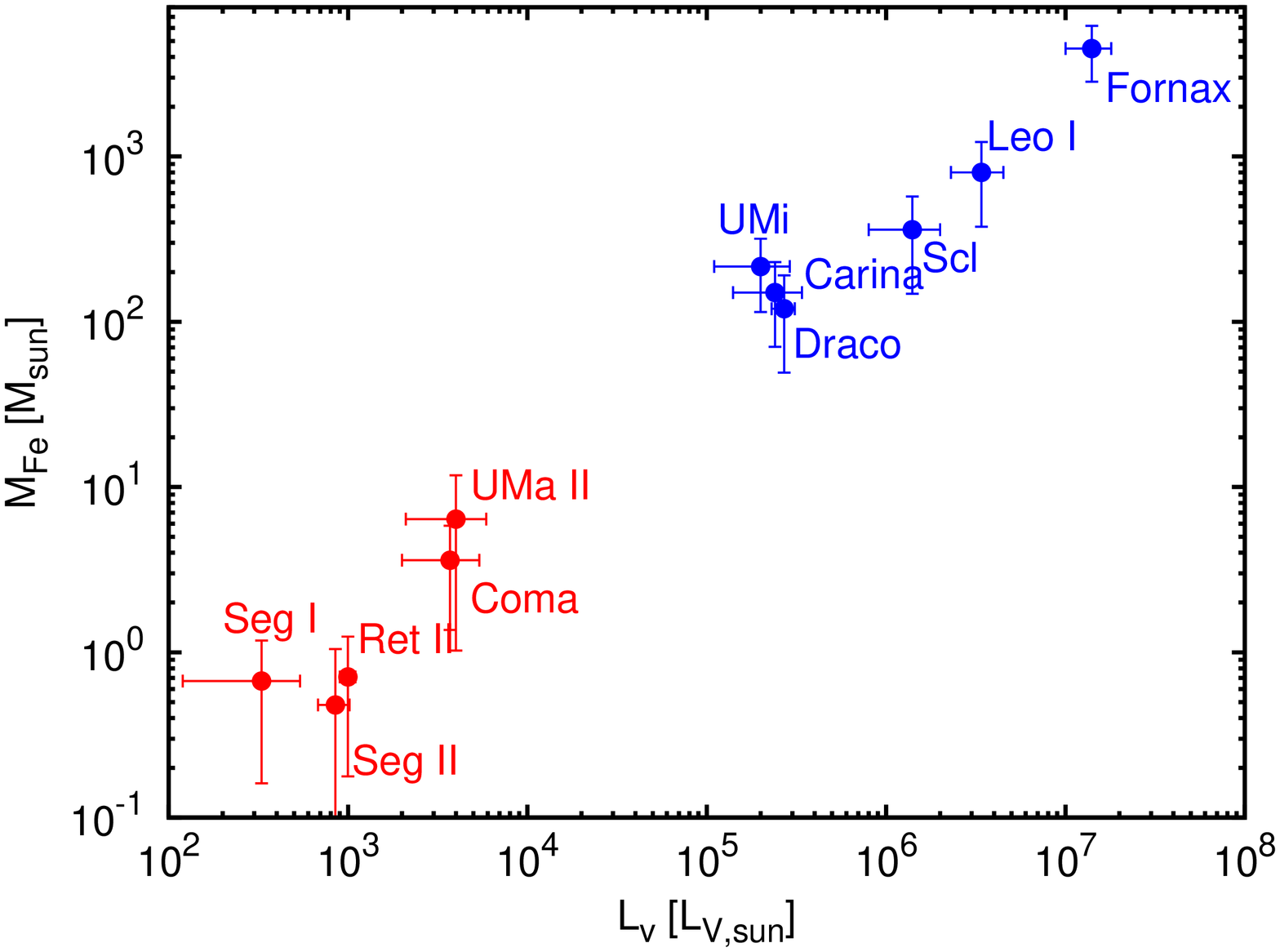}
\includegraphics[angle = 270, bb=50 50 554 770,width=90mm]{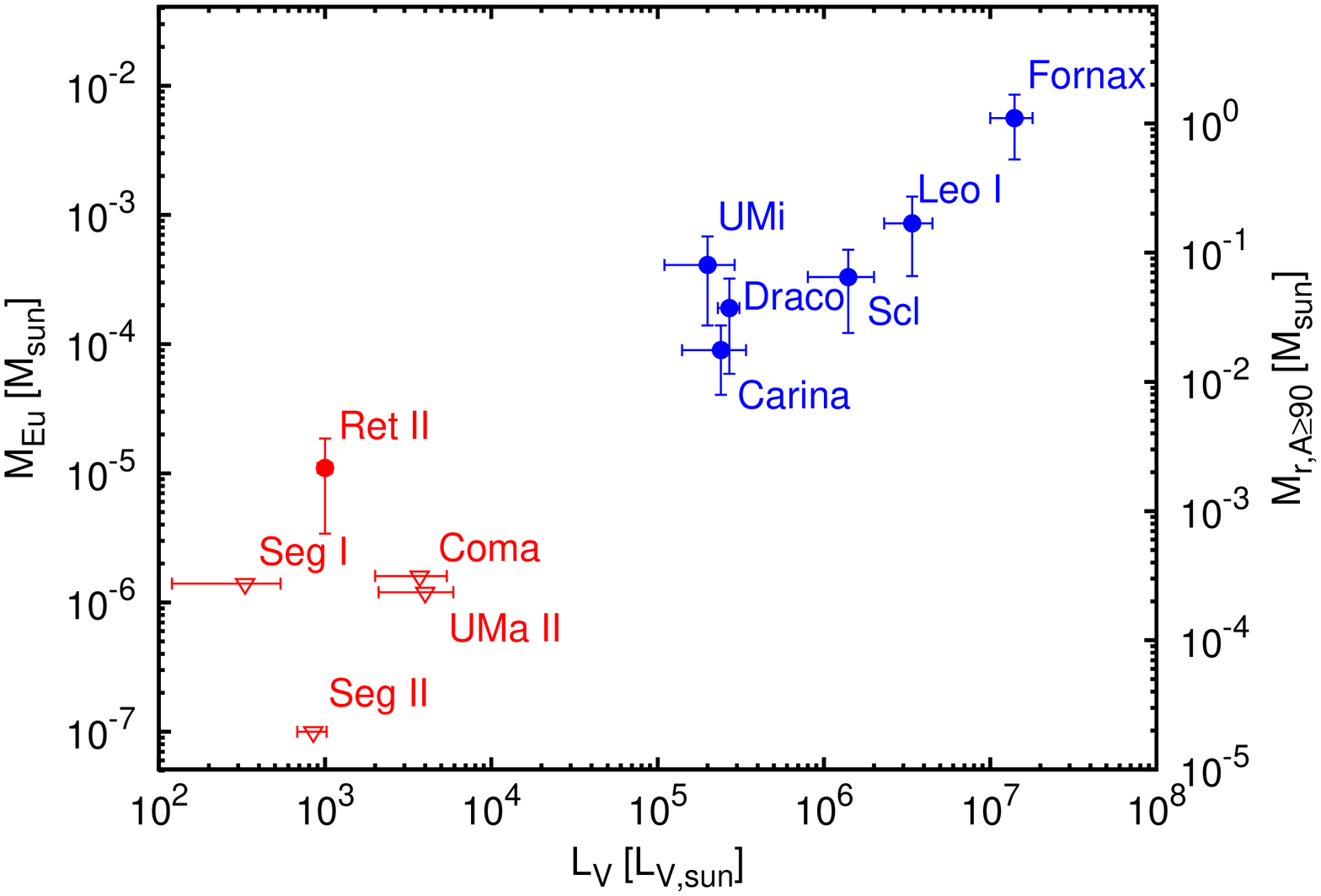}
\caption{
 Fe~(left) and Eu~(right)  masses in dwarf galaxies. The horizontal axis
shows the total V-band luminosity of each galaxy. The right vertical axis  of the left panel 
shows the mass of $r$-process elements with $A\geq 90$ estimated by Eq.~(\ref{rmass}).
Blue and red points depict the classical dwarfs and UFDs respectively. Open triangles 
show upper limits on the masses.}
\label{fig:eu}
\end{figure}

Figure~\ref{fig:eu} depicts the estimated total Fe and Eu masses for each dwarf galaxy as a function
of the total stellar luminosity. Note that the Fe mass is proportional to the luminosity within uncertainties,
implying that the number of SNe producing Fe is proportional to the number of stars. For the Eu masses,
while a similar trend can be found, the dispersion is larger for the smaller galaxies.
We find $M_{\rm Eu}\approx 10^{-5}M_{\odot}$ for Reticulum II. For the other UFDs there are strong upper limits, 
in some cases as small as $10^{-7}M_{\odot}$. This led \cite{ji2016Nature} to suggest that there was only a single 
$r$-process production event in Reticulum II and none in the other UFDs.

\begin{table*}[ht]
\begin{center}
\small
%\begin{minipage}{18cm}
\caption[]{
  Summary of derived quantities for the dwarf galaxies in the sample.\\
  $^{a}$ the initial gas mass estimated with the assumption of constant velocity dispersion up to $r_{90}$.\\
  $^{b}$ the initial gas mass estimated with the NFW model.
}
%\hspace{-2.2cm}
\scalebox{0.9}{
 \begin{tabular}{ccccc} \hline \hline
Type & Object  & $r_{90}$~[pc]& $M_{g}(\!<\!r_{90})^{a}~[10^{6}M_{\odot}]$ & $M_{g}(\!<\!r_{90})^{b}~[10^{6}M_{\odot}]$ \\ \hline 
\multirow{5}{*}{UFDs} & Reticulum II  & $120^{+7}_{-4}$ &  $0.28^{+0.15}_{-0.12}$
& -\\
& Segue I & $110\pm 3$ & $0.37\pm 0.21$ 
& -  \\
& Segue II & $130\pm 20$ &  $0.27\pm 0.28$
& - \\
& Ursa Major II  & $520\pm 90$ &  $4.2\pm 1.8$ 
& - \\
& Coma Berenices & $290\pm 40$ & $1.1\pm 0.8$ 
& - \\ 
\hline
\multirow{6}{*}{other dwarfs} 
& Draco  & $730\pm 50$ &  $11\pm 3$
& $9.8^{+3.5}_{-2.7}$\\
& Leo I  & $920\pm 70$ & $14 \pm 4$
& $13\pm 4$\\
& Fornax  & $2500\pm 100$ & $61\pm 10$
& $30\pm 4$\\
%Sgr  & $5800\pm 190$ & $11.4\pm 0.7$ & $81\pm 13$ & - \\
& Carina  & $900\pm 90$ &$7.0\pm 2.7$
& $5.3 \pm 1.5$\\
& Ursa Minor  & $1000\pm 200$ &  $17\pm 5$
& $14\pm 6$\\
& Sculptor  & $970\pm 150$ & $15\pm 4$
& $13\pm 5$\\
\hline \hline\\
\label{tbl:mass}
\end{tabular}
}
\end{center}
%\end{minipage}
\end{table*}

\section{The $r$-process and SNe event rate and amount of mass produced per event} 
\label{sec:like}

\subsection{The statistical method}
\label{stat}
We aim to constrain the event rate and mass per event for both SNe and $r$-process events.
We begin with a description of the general statistical method that we use in both cases.
We define the likelihood function in terms of two unknown parameters: $R$, the average 
rate of events, and $\tilde{m}_X$, the mass of an element $X$ produced by a single event.

Using Poisson statistics, we calculate the probability of obtaining $N$ events in the $i$th galaxy, given an expected number of events $N_{R,i}$.
We assume that the mass of $X$ produced per event follows a normal distribution with an average of $\tilde{m}_X$ and 
a standard deviation of $0.5 \tilde{m}_X$. For each $N$ the associated probability is then the Poisson 
probability of obtaining $N$ events times the probability that $N$ events would lead to an accumulated mass of element $X$ in the $i$th galaxy,
$M_{X,i}$ (calculated from the observed values of [X/H] and $M_g$). Summing these 
probabilities over all $N$ values, we obtain the final probability: $P_i(M_{\rm X,i} |R,\tilde{m}_X)$. 
Finally, we write the likelihood function as
\begin{equation}
 L(R,M_X)=\Pi_i P_i(M_X |R,\tilde{m}_X) \ .
\end{equation}
We maximize, of course, this function over its free parameters.

\subsection{The SN rate and iron production in dwarf galaxies}
\label{sec:Fe}
\begin{figure}[h]
\centering
\includegraphics[scale=0.3]{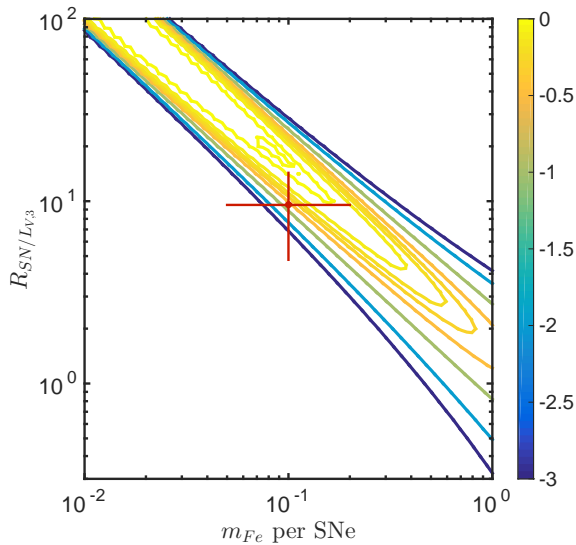}
\caption{A likelihood analysis for Fe production as a function of the Fe mass per event $\tilde{m}_{\rm Fe}$ and the
number of SNe per $10^3L_{V,\odot}$ stellar luminosity (roughly $10^3$ stars) for the entire sample of dwarf galaxies.
Also shown as a cross is the Fe mass per ccSN and ccSN rate in the local Universe~\citep{Drout2011,Li2011,horiuchi2011ApJ,Kushnir2015}.
Here we adopt an uncertainty in the mass and rate by a factor of $2$.}
\label{fig:likelihoodSNe}
\end{figure}

We begin applying the method described in \S \ref{stat} to constrain the rate of Fe-producing SNe and the amount of Fe produced per event in these dwarf galaxies.
The resulting likelihood function is seen in Fig. \ref{fig:likelihoodSNe}. The main constraint is on the product of the two parameters. This reflects the fact that the average Fe mass per solar luminosity should match the observed ratio. The actual peak of the likelihood function is not strongly constrained, although interestingly, it coincides with an Fe mass per event of $\tilde{m}_{Fe}=0.1M_{\odot}$ and a rate of $\approx 16$ SNe per $10^3L_{V,\odot}$ (approximately $10^3$ stars).

These values are in fact expected from SNe studies.
In the local Universe, two thirds of ccSNe are type II supernovae that typically 
produce $0.02M_{\odot}$ of Fe and type Ibc that produce $0.2M_{\odot}$~\citep{Drout2011,Kushnir2015}, and 
one third of ccSNe are type Ibc~\citep{Li2011}.
Therefore, the Fe production is dominated by type Ibc supernovae. 
In addition, since the star formation epoch in dwarf galaxies had lasted for a short time-sclae of $\sim 1$~Gyr,
we assume that type Ia SNe don't contribute to the Fe production. Thus, $0.1M_{\odot}$ per ccSN is expected for dwarf galaxies. In addition, the rate of $16$ SNe per $10^3L_{V,\odot}$ is consistent with observations suggesting $1$ SN per 100$M_{\odot}$ of star formation (e.g. \citealt{Li2011,horiuchi2011ApJ}). The observational constraints on the rate and Fe mass per event from the local Universe are depicted by the red cross
in Fig. \ref{fig:likelihoodSNe}.
This result suggests that the Fe production in each ccSN and their rate in metal-poor environments 
are similar to those in the local Universe. In addition, it demonstrates the validity of this method.  

\subsection{Results and implications for $r$-process production in dwarf galaxies}
We turn now to the production of Eu in $r$-process events. We calculate the rates relative to the ccSN rate, $R_{rp/SN}$.
Using the results of the previous section (and independent analyses, see \S \ref{sec:Fe}), we assume an average production of $0.1M_{\odot}$ of Fe per ccSN.
The average number of $r$-process events in the $i$th galaxy is then $N_{R,i}=R_{rp/SN}\times (M_{\rm Fe,i}/0.1M_{\odot})$, where $M_{\rm Fe,i}$ is the total amount of Fe in the same galaxy.

\begin{figure}[h]
\includegraphics[scale=0.3]{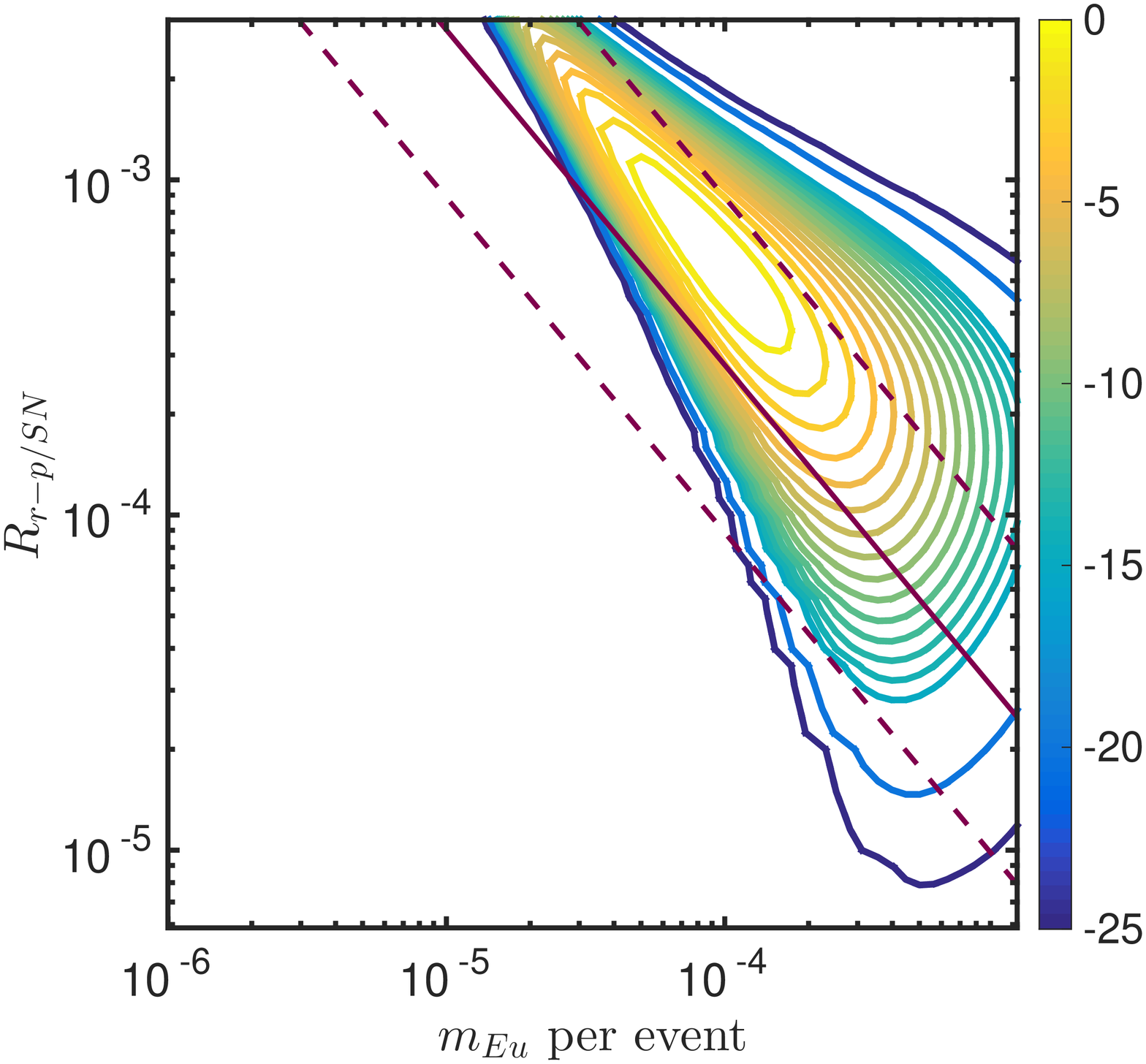}
\includegraphics[scale=0.3]{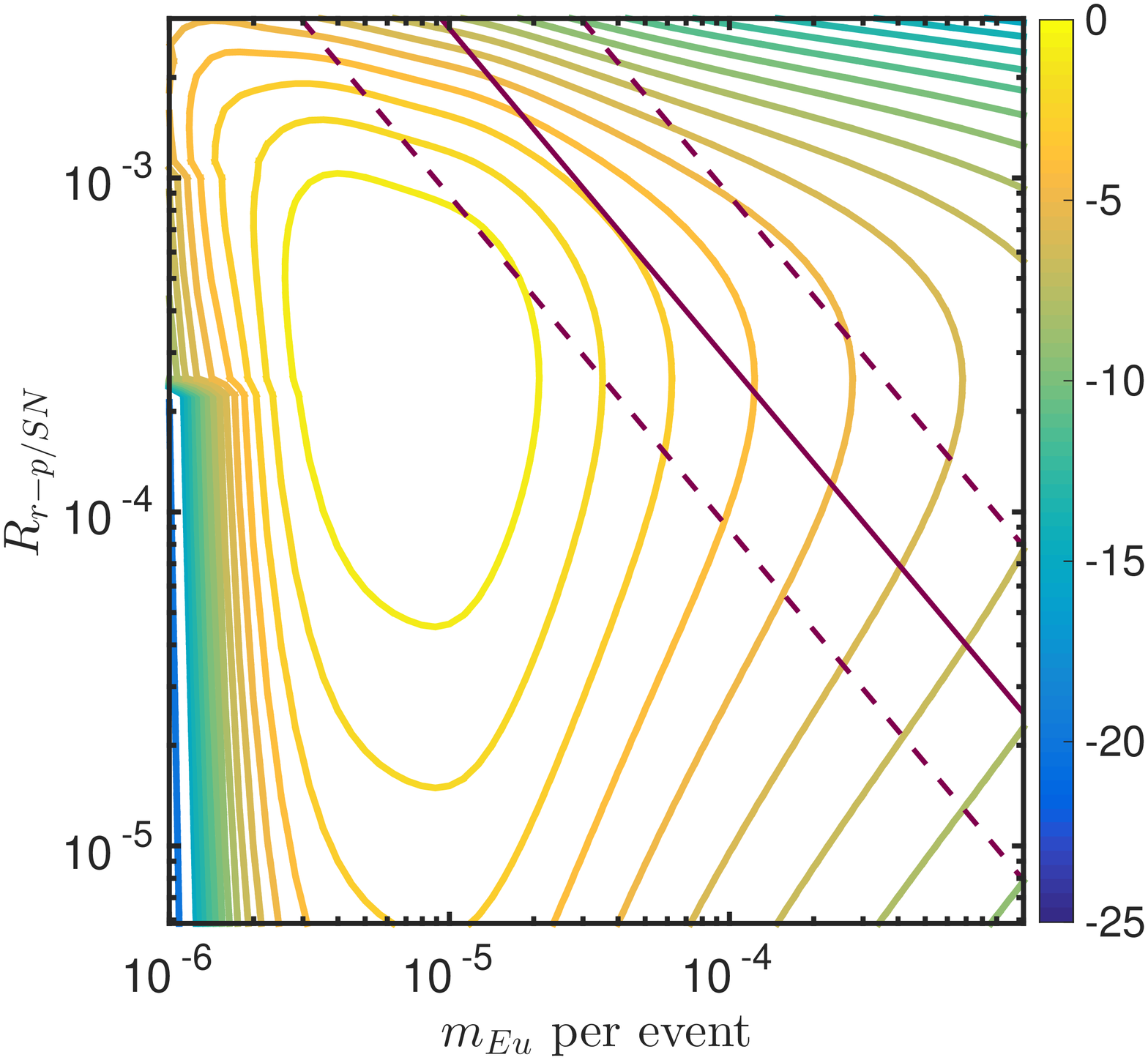}
\caption{
A likelihood analysis for $r$-process production as a function of the Eu mass per event $\tilde{m}_{\rm Eu}$ and the
relative $r$-process to ccSN event rate, $R_{rp/SN}$, for the entire sample of dwarf galaxies (left)
and for UFDs only (right). Also overlaid is the permitted range (between the dashed lines) and the best fit relation (solid line) inferred from Milky Way observations.
}
\label{fig:likelihood}
\end{figure}

Figure \ref{fig:likelihood} shows the results for two cases:  all dwarf galaxies~(left) and only UFDs~(right).
The likelihood function peaks at a rate relative to the SN rate of $R_{rp/SN}=6\times 10^{-4}$ ($R_{rp/SN}=3\times 10^{-4}$), and at a mass per event of $\tilde{m}_{\rm Eu}=9\times 10^{-5}M_{\odot}$ ($\tilde{m}_{\rm Eu}=7\times 10^{-6}M_{\odot}$), corresponding to a total mass of $r$-process matter produced per event of roughly $\sim 1.7 \times 10^{-2}M_{\odot}$ ($1.3 \times 10^{-3}M_{\odot}$). The strong upper limits on the Eu abundance of some galaxies, coupled with the detection of a significant amount of Eu in galaxies of similar masses, implies that the Eu mass produced per event should be approximately the smallest amount detected in any given galaxy.
Notice that this situation is very different than the case for Fe production, where the same statistical method results in no preference for a small number of events per galaxy.
This leads to a relative rate of $r$-process to ccSN events which is $\ll 1$. Thus we can rule out the possibility that normal ccSNe dominate the production of $r$-process materials in dwarf spheroidal galaxies.

Interestingly, this rate of Eu production is consistent with the constraints from our own, much more massive Galaxy.
Assuming one ccSN per 100$M_{\odot}$ of star formation,
and given a total stellar mass of $6.4\times 10^{10}M_{\odot}$ \citep{mcMillan2011MNRAS},
we obtain $6.4\times 10^8$ such SNe in the history of the Milky Way \citep{venn2004AJ}.
Combining this with a total Eu mass of $\approx 18M_{\odot}$, 
%which is obtained by using $[\langle {\rm Eu/H}\rangle]\approx -0.15$ for the thin disk and $[\langle {\rm Eu/H}\rangle]\approx -0.2$ 
%for the thick disk \citep{venn2004AJ} and assuming $[\langle {\rm Eu/H}\rangle]\approx 1$ for the bulge, 
leads to:
$R_{rp/SN}=2.8\times 10^{-3} (\tilde{m}_{\rm Eu}/10^{-5}M_{\odot})^{-1}$ with a systematic uncertainty of a factor of $\sim 3$.
The range permitted by these observations is depicted as the area between the dashed
lines in Fig. \ref{fig:likelihood}. As seen in that figure, this range is consistent with
the best fit parameters found for the dwarf galaxies. 
Furthermore, the low rate is consistent with $R_{rp/SN}<5\cdot 10^{-3}$ which \cite{hotokezaka2015NaturePhys} find in order to 
reproduce the $^{244}$Pu abundances of both the early solar system material~\citep{turner2007E&PSL} and the present-day deep-sea archives~\citep{Wallner2015NatCo}.
These suggest that the $r$-process production in the Milky Way is dominated by the same events as in dwarf galaxies. Therefore, these results for the rate and mass per event can be generalized, and the same process could be the dominant source of $r$-process production in the whole Universe.

\subsection{The robustness of the method and dependence on parameters}
We turn now to exploring the dependence of the results on the assumed halo masses.
As mentioned in \S \ref{sec:initial}, there is some uncertainty in the values of these masses. We repeat the calculation (for all dwarf galaxies)  with halo masses of
$M_h(<r_{50})$ (smaller by about an order of magnitude as compared with $M_h(<r_{90})$) and $3M_h(<r_{90})$. For $M_h(<r_{50})$ ($3M_h(<r_{90})$), the likelihood peaks at an Eu mass per event of $\tilde{m}_{\rm Eu}=3\times 10^{-5}M_{\odot}$ ($\tilde{m}_{\rm Eu}=2.5\times 10^{-4}M_{\odot}$) and at a rate relative to the SN rate of $R_{rp/SN}=2\times 10^{-3}$ ($R_{rp/SN}=3\times 10^{-4}$). As expected, the amount of Eu mass per event is approximately the minimal Eu mass observed in any galaxy and thus tracks linearly the halo mass. In addition, since $R_{rp/SN}$ is roughly
\begin{equation}
R_{rp/SN}=\frac{M_{\rm Eu} \tilde{m}_{\rm Fe}}{M_{\rm Fe} \tilde{m}_{\rm Eu}}
\end{equation} 
and since $M_{Eu},M_{Fe},\tilde{m}_{\rm Eu}$ are linear in the halo mass, it follows that $R_{rp/SN}$ is inversely proportional to this quantity.
Most importantly, the conclusion that $R_{rp/SN} \ll 1$ holds in all of these cases.

\section{Implications to the scenarios of $r$-process production}

\subsection{Double neutron star mergers}
\label{sec:DNSRates}
The $r$-process mass per event that we find in the likelihood analysis described in \S \ref{sec:like}, is roughly consistent with expectations for DNS mergers from simulations.
These simulations typically find ejecta masses of $10^{-3}$--$10^{-2}M_{\odot}$
corresponding to Eu masses of $5\times 10^{-6}$--$5\times 10^{-5}M_{\odot}$
\citep{korobkin2012MNRAS,bauswein2013ApJ,perego2014MNRAS,just2015MNRAS,sekiguchi2015PRD,radice2016MNRAS,wu2016}.
The results are also consistent with macronova estimates~\citep{berger2013ApJ,hotokezaka2013ApJ,piran2014,jin2016}.

The strong upper limits of $\lesssim 10^{-6}M_{\odot}$ for the other UFDs imply that  such events didn't take place in
these UFDs. We turn now to estimating the expected rates of DNS mergers in these galaxies and comparing them to the results from the likelihood analysis in \S \ref{sec:like}.
The expected number of $r$-process production events  by DNS inside a UFD during its star formation epoch is 
\begin{eqnarray}
\label{general_rate}
R_{rp/SN} = f(v_{CM}<v_{\rm esc}, t_{\rm m}<t_{\rm SF})
\frac{\rm Number~of~mergers}{\rm Number~of~ccSNe},
\end{eqnarray}
where $f(v_{CM}<v_{\rm esc}, t_{\rm m}<t_{\rm SF})$ is the fraction of mergers with a  center of mass velocity, $v_{CM}$ smaller than the escape velocity of a UFD, $v_{\rm esc}$, and with a merger time, $t_{\rm m}$, shorter than the duration of the star formation phase, $t_{\rm SF}$. 
The small escape velocities from UFDs (down to $\approx 15~{\rm km/s}$ in some cases)
and the relatively short durations of their star formation, $\lesssim 1$~Gyr, reduce the number of
mergers providing $r$-process elements in the star forming gas.
However, in \cite{beniamini2016} we have shown that since many DNS systems receive small kicks at birth \citep{BP(2016)}, a
large fraction of them would have $v_{CM}<15 $\,km\,s$^{-1}$.
Furthermore, given limits on the separation of the double pulsar system before its second collapse, a significant fraction of systems
are expected to both remain confined in UFDs and merge within less than a Gyr.
All together, we estimate that: $6\times 10^{-2} \lesssim f(v_{CM}<v_{\rm esc}, t_{\rm m}<t_{\rm SF}) \lesssim 0.6$.
Therefore, using our result in \S ~\ref{sec:like}, the number of mergers per ccSN is estimated as $5\times10^{-4}$---$ 2\times 10^{-2}$.

We compare the  results also with the merger rate estimated using
short GRB observations.
The local short GRB rate is estimated as
$4.1^{+2.3}_{-1.9}f_{b}^{-1}~{\rm Gpc^{-3}\,yr^{-1}}$~\citep{wanderman2015MNRAS},
where $f_b^{-1}$ is a beaming factor that is in the range $1<f_b^{-1}\lesssim 100$~\citep{Fong2015ApJ}.
The rate of local ccSNe is $7.1_{-1.3}^{+1.4}\times 10^{4}~{\rm Gpc^{-3}\,yr^{-1}}$.
Therefore, the number of short GRBs per ccSN is $5.8^{+3.2}_{-2.8}\times 10^{-5}f_{b}^{-1}$,  consistent with our estimates above.

\subsection{Peculiar core-collapse supernovae}
\label{sec:ccSNe}
The low rate and large amount  of $r$-process elements per event $\sim 0.01M_{\odot}$
challenge the standard ccSN scenarios. A neutrino-driven wind of proto-neutron stars of  normal ccSNe is clearly incompatible with the low rate.  
However, the low rate can be consistent with a  peculiar astrophysical
explosion associated with a core collapse of massive stars. 
For instance, a magnetically powered outflow from newly born 
magnetars has been proposed as the site of the $r$-process nucleosynthesis 
\citep{suzuki2005ApJ,metzger2008ApJ,Winteler12,vlasov2014MNRAS,nishimura2015ApJ,Moesta15}.
This could happen in a peculiar SN \citep[e.g.][]{Winteler12} or in a 
long GRB, in which case the magnetic outflow also powers the GRBs themselves.
A neutrino driven-wind from an accretion torus of newly formed black holes~\citep{wanajo2012ApJ}
and fall-back supernovae~\citep{fryer2006ApJ} have also been suggested as the production sites.

The rate of long GRBs has been estimated as $1.3^{+0.6}_{-0.7}$~Gpc$^{-3}$ yr$^{-1}$ before beaming corrections \citep{Wanderman2010}. 
This is comparable, up to a factor of 3 with the rate of short GRBs. 
Given a lower rate, more $r$-process mass per event is required. 
But, of course, this also depends on an unknown beaming factor.   
Low luminosity GRBs and superluminous SNe 
have  also been proposed to involve magnetized outflows (or for the latter, winds from accretion tori around black holes). 
The rates of these events 
are  $230^{+450}_{-190}$~Gpc$^{-3}$ yr$^{-1}$~\citep{soderberg2006} and $91^{+76}_{-36}$~Gpc$^{-3}$ yr$^{-1}$~\citep{prais2016MNRAS},
corresponding to a rate relative to ccSNe of approximately $10^{-3}-10^{-2.5}$. This is at the upper end of the permitted range that we find (\S \ref{sec:like}) and would, in turn, require the $r$-process mass per event to be at the low end of the permitted range ($\approx 6\times 10^{-3}M_{\odot}$).

\cite{wanajo2012ApJ} have estimated
that a neutrino-driven wind from a black-hole accretion torus can produce $\sim 10^{-3}M_{\odot}$, 
which seems too small to be consistent with the permitted range that we find 
here\footnote{The amount of the torus wind depends on the black-hole spin~\citep{just2015MNRAS,fernandez2015MNRAS}. There may be parameter regions where more massive winds can be produced.}. 
Other estimates of $r$-process material production yield 
 $\approx 0.01 M_\odot$ for a  SN magnetized driven jet
 \citep{nishimura2015ApJ} and $\lesssim 0.01M_{\odot}$  for a  magnetized neutrino-driven wind   \citep{vlasov2014MNRAS}.  Since these processes require rapid rotation, and hence low metallicity, these authors
estimate that the rate of such events is low compared to regular SNe.
However, the exact rates are uncertain. 

\section{Conclusions and Discussions}
\label{sec:conclusions}
We have explored the question of iron and $r$-process production in dwarf spheroidal galaxies in a model independent fashion.
We assume that all dwarf galaxies (or alternatively, only UFDs) share the same mechanism for production of these elements.
We then use the element abundances and apply a likelihood method to estimate the most likely rates and element mass produced per such event.
Using the mass within the radius that contains $90\%$ of the galaxies' light, $M_h(<r_{90})$, as a proxy for the halo masses (which in turn provides the initial gas mass), we find that on average $0.1M_{\odot}$ of Fe is produced per ccSN and that the rate is approximately 16 ccSNe per
$10^3L_{\odot}$, verifying the estimated values of these parameters from independent studies using different methods \citep{horiuchi2011ApJ}.

Turning now to $r$-process events, we find that the Eu mass per event is between $3\times 10^{-5}M_{\odot}<\tilde{m}_{\rm Eu}<2\times 10^{-4}M_{\odot}$ ($2.5\times 10^{-6}M_{\odot}<\tilde{m}_{\rm Eu}<3.5\times 10^{-5}M_{\odot}$ for UFDs only).
The very strong upper limits on the Eu mass in some galaxies combined with detections of significantly larger masses in other dwarfs,
imply that Eu production is dominated by rare events and that the minimal Eu mass observed in any galaxy
is approximately the amount of Eu mass produced per event.
This corresponds to a total $r$-process mass of $6\times 10^{-3}M_{\odot}<\tilde{m}_{r-process}<4\times 10^{-2}M_{\odot}$ ($5\times 10^{-4}M_{\odot}<\tilde{m}_{r-process}<7\times 10^{-3}M_{\odot}$ for UFDs only), assuming that $r$-process events dominate the production of elements with mass number $A\geq90$. The rate of these events is $2.5\times 10^{-4}<R_{rp/SN}<1.4\times 10^{-3}$ as compared with the ccSN rate ($1.6\times 10^{-5}<R_{rp/SN}<1.4\times 10^{-3}$ for UFDs only).
We conclusively rule out the possibility that the rates of $r$-process events could be comparable to that of ccSNe.
Interestingly, both the $r$-process mass per event and the rate are, however, consistent with expectations for DNS mergers. However, rare magnetic driven SNe, long GRBs, superluminos SNe or even low luminosity GRBs are all possible candidates. 
Furthermore, these values are consistent with the total Eu mass observed in our own Galaxy. This suggests that the same mechanism
is behind the production of $r$-process events in both UFDs and the Milky Way, and that it may be the dominant mechanism for production of $r$-process elements as a whole in the Universe. Association of a gravitational-wave event from a DNS merger and a clear macronovae from which the amount of $r$-process production can be well estimated could therefore confirm or reject this scenario. 

A large scatter in the abundances of $r$-process elements is also seen in metal poor stars in our own Galaxy, which again supports rare events (e.g., \citealt{tsujimoto2014A&A,Cescutti2014,wehmeyer2015MNRAS}).
In addition, \cite{hotokezaka2015NaturePhys} have shown that the rarity of $r$-process events is
broadly consistent with  $^{244}$Pu abundances of both the early solar
system material~\citep{turner2007E&PSL} and the present-day deep-sea archives~\citep{Wallner2015NatCo}, which are difficult to explain otherwise.
The $r$-process element enrichment in 
halo stars with very low metallicities, proposed as a problem for this scenario \citep{argast2004A&A}, 
can be resolved  by taking into account
the turbulent mixing of material in the galaxy~\citep{piran2014}, the  assembly of
sub-halos during the formation of the Galaxy~\citep{komiya2014ApJ,vandevoort2015MNRAS,ishimaru2015ApJ,shen2015ApJ},
the large typical distances between the locations of the stellar collapses and the eventual mergers \citep{beniamini2016} as well as
the likely possibilities of rapid mergers and stellar collapses with small amounts of mass ejecta \citep{beniamini2016,BP(2016)}.

Given the orbits of the observed DNS systems in the Galaxy, a significant amount of DNS systems are expected to both remain confined within UFDs and merge within a relatively short time since formation \citep{beniamini2016}. Using these results and assuming that $r$-process production is dominated by DNS mergers, we estimate the DNS merger rate to be $5\times10^{-4}$--$ 2\times 10^{-3}$ per ccSN. This is comparable to the rates estimated using other methods such as from short GRBs or from population synthesis. 

There are two major differences between the DNS and core-collapse scenarios.
First, DNS mergers have delay times  relative to the star formation but the core-collapse events don't.  Moreover since  the rare core-collapse events, discussed above, may take place preferably 
in low-metallicity environments, they may even precede the average star formation and produce $r$-process elements  in very low metallicity environments. However, the region will be ``polluted" by mass outflow from the progenitor stars, Oxygen and Carbon from winds and Fe from the associated SN.    Second, during the formation of the second neutron star in a DNS, the center of mass receives a velocity of  $\sim 15$~km/s~\citep{beniamini2016}. This 
is slower than the typical escape velocity from UFDs but fast enough to run away 
from the star forming regions where they have been formed. Thus the mergers can take place in low-metallicity environments that are ``unpolluted" by the progenitors.  This depends, of course, on the pace of turbulent mixing within the UFD.
These differences should be reflected 
in the distribution of Eu and Fe within the galaxies and in the chemical history of the Galaxy.   Abundance measurements for a larger   number of stars in Reticulum 2 and other  dwarf 
galaxies may provide vital information that may enable us to distinguish between the two  scenarios.

We thank Alex Ji and Kohei Inayoshi for useful discussions. 
This work was supported in part by an Israel Space Agency (SELA) grant, the Templeton Foundation and the I-Core center for excellence ``Origins" of the ISF.

%\bibliography{Rprocess}
\end{document}